\documentclass[pre,groupedaddress,superscriptaddress,reprint,twocolumn,showpacs,notitlepage]{revtex4}

\usepackage{amsfonts}
\usepackage{amsmath}
\usepackage{amssymb}
\usepackage{graphicx}
\usepackage{hyperref}
\usepackage{color}
\usepackage{mathrsfs}
\usepackage{bbm}
\usepackage{subfigure}
\usepackage{times,txfonts}

\newcommand{\ignore}[1]{}
\newcommand{\nobibentry}[1]{{\let\nocite\ignore\bibentry{#1}}}
\newcommand{\bibfnamefont}[1]{#1}
\newcommand{\bibnamefont}[1]{#1}

\newcommand{\ket}[1]{\left\vert#1\right\rangle}
\newcommand{\bra}[1]{\left\langle#1\right\vert}
\begin{document}
%\nobibliography*

\title{Multi-stage quantum absorption heat pumps}

\author{Luis A. Correa}
\email{lacorrea@ull.es}
\affiliation{School of Mathematical Sciences, The University of Nottingham, University Park, Nottingham NG7 2RD, UK}
\affiliation{IUdEA Instituto Universitario de Estudios Avanzados, Universidad de La Laguna, 38203 Spain}
\affiliation{Dpto. F\'{\i}sica Fundamental, Experimental, Electr\'{o}nica y
Sistemas, Universidad de La Laguna, 38203 Spain}

\date\today

\begin{abstract}
It is well known that heat pumps, while being all limited by the same basic thermodynamic laws, may find realization on systems as `small' and `quantum' as a three-level maser. In order to quantitatively assess how the performance of these devices scales with their \textit{size}, we design generalized $N$-dimensional ideal heat pumps by merging $N-2$ elementary three-level stages. We set them to operate in the \textit{absorption chiller} mode between given hot and cold baths, and study their maximum achievable cooling power and the corresponding efficiency as a function of $N$. While the efficiency at maximum power is roughly size-independent, the power itself slightly increases with the dimension, quickly saturating to a constant. Thus, interestingly, scaling up autonomous quantum heat pumps does not render a significant enhancement beyond the optimal double-stage configuration.
\end{abstract}

\maketitle

\section{Introduction}

An absorption heat pump is a multi-purpose thermal device in which the driving force is mostly heat, rather than mechanical work \cite{gordon2000cool}. The first design thereof was patented by Ferdinand Carr\'{e} in 1860 and played a prominent role in the early development of refrigerators. Anecdotally, in 1927, Albert Einstein and Le\'{o} Szil\'{a}rd presented an absorption fridge running \textit{solely} on heat, that operated at nearly constant pressure, requiring no moving parts \cite{einstein1927refrigeration}. In spite of the obvious advantages of absorption technologies in terms of safety, reduced environmental impact or better exploitation of the freely available thermal resources, mechanical compression cycles are preferred for most industrial and commercial applications, as these are usually much more efficient. Consequently, considerable effort has been devoted over the last decades to improve the performance of absorption heat pumps \cite{srikhirin2001review} in order to make them compete with their power-driven analogues. For instance, one could think of scaling up absorption systems into larger \textit{multi-stage} configurations capable of reusing the residual output heat for further cooling, thus boosting their overall performance \cite{ziegler1993multi,lawson1982water,devault1990ammonia,grossman1995simulating}. 

It was already acknowledged in the late 1950s that a three level \textit{maser} selectively coupled to two heat baths and a coherent driving field, is a valid embodiment for a quantum heat engine \cite{PhysRevLett.2.262,5123190,PhysRev.156.343}, which running in reverse ultimately amounts to a power-driven fridge. Ever since, quantum heat engines and refrigerators have been object of extensive study \cite{1310.0683v1,PhysRevE.76.031105,PhysRevE.79.041129,geva1996quantum,PhysRevE.87.012140,PhysRevE.81.051129,PhysRevE.82.011120,PhysRevLett.106.070401,PhysRevA.87.063845,gelbwaser2013autonomous}, including detailed experimental proposals \cite{PhysRevLett.109.203006,1308.5935v1}, and the development of strategies to overcome their ultimate thermodynamics limitations \cite{PhysRevE.86.051105,1303.6558v1,PhysRevA.87.063845,gelbwaser2013autonomous}.

On the other hand, a three-level maser operating between heat baths only (i.e. without driving) realizes a \textit{quantum} absorption heat pump  \cite{PhysRevE.64.056130}, that can function either as a fridge or as a `heat transformer'. In particular, quantum absorption fridges have attracted a lot of attention \cite{PhysRevLett.108.070604,PhysRevLett.105.130401,1751-8121-44-49-492002,PhysRevE.85.051117,PhysRevE.85.061126,1305.6009v1,PhysRevLett.108.120602,gelbwaser2013autonomous}, mostly because of their inherent simplicity, that allows to gain insight into the emergence of the laws of thermodynamics from individual open quantum systems \cite{e15062100}. 

In many practical situations, absorption chillers might also be a preferable alternative to power-driven quantum devices. Nonetheless, as in the classical case, these are usually much less efficient and powerful than the best reversed heat engines \cite{1308.4174}, which turns the optimization of their performance into a matter of paramount importance. Some steps have been already undertaken in this direction, by identifying the upper bounds on a suitable figure of merit for the cooling performance, and the design prescriptions for their saturation \cite{PhysRevE.87.042131}, and by proposing reservoir engineering techniques \cite{PhysRevA.57.548,shahmoon2013engineering} in order to \textit{push} those bounds further \cite{1308.4174}.

Notwithstanding the important differences between classical and quantum absorption heat pumps, it is most natural to ask whether significant enhancement can be expected from scaling up quantum thermodynamic cycles into ``larger'' multi-stage networks. In order to quantitatively assess the size scaling on the performance of heat pumps, we constructively build generalized $N$-level ideal absorption devices by coupling together $N-2$ elementary three-level cycles, somewhat in the spirit of \textit{parallel} multi-stage classical absorption devices \cite{gordon2000cool}. We then compute the efficiency at maximum cooling power and the maximum power itself as a function of $N$ for $3\leq N\leq 10$. While the efficiency proves to be roughly size-independent, the corresponding power only increases mildly with the number of stages, quickly saturating to a constant asymptotic value. Therefore, and again in total correspondence with the classical scenario, double-stage absorption heat pumps appear as the most reasonable compromise between simplicity and performance. Interestingly, this case of $N=4$ happens to correspond with the two-qubit model recently introduced in \cite{PhysRevLett.108.070604}, thus marking it as a target for experimental implementations.

We also carry out an extensive numerical investigation that supports the general validity and tightness of the model-independent performance bound established in \cite{1308.4174}, for all ideal multi-stage absorption refrigerators. Finally, we compare the non-ideal eight-level absorption fridge of \cite{PhysRevLett.105.130401} with all its ideal counterparts, when operating under the same conditions. We find that ideal fridges largely outperform non-ideal ones, not only when it comes to their maximum achievable efficiency but, especially, in terms of their maximum cooling power.

This paper is structured as follows: We start by providing a brief review of quantum absorption heat pumps in Sec.~\ref{review}. In Sec.~\ref{system}, we introduce our constructive scheme to build generalized multi-stage parallel cycles, discuss their microscopic model and corresponding master equation, and obtain the steady-state heat currents flowing across the system. We shall be then, in Sec.~\ref{results}, in the position to quantitatively compare the maximum cooling power and the corresponding efficiency of devices with increasing $N$ and fixed input thermal resources. Finally, in Sec.~\ref{conclusions}, we summarize and draw our conclusions.
\newline

\section{Ideal quantum absorption heat pumps \label{review}}

Let us consider a three-level system with Hamiltonian $\hat H_3 = \hbar\omega_c \ket{2} + \hbar\omega_h \ket{3}$ in its energy eigenbasis $\{\ket{1},\ket{2},\ket{3}\}$. We shall allow for a \textit{very weak} dissipative interaction between the transition $\ket{1}\leftrightarrow\ket{2}$ and a `cold' heat bath at temperature $T_c$. Likewise, we couple the transitions $\ket{2}\leftrightarrow\ket{3}$ and $\ket{3}\leftrightarrow\ket{1}$ very weakly to a high temperature `work' reservoir and a `hot' bath respectively, such that $T_w>T_h>T_c$. Once its asymptotic state builds up, this system operates as the most fundamental quantum absorption heat pump \cite{PhysRevE.64.056130}, relying on the imbalance between the steady-state rates associated with the cooling cycle $\ket{1}\xrightarrow{c}\ket{2}\xrightarrow{w}\ket{3}\xrightarrow{h}\ket{1}$, and its complement, the heat-transforming cycle $\ket{1}\xrightarrow{h}\ket{3}\xrightarrow{w}\ket{2}\xrightarrow{c}\ket{1}$. 

When the asymptotic rate of the cooling cycle exceeds that of the heat-transforming cycle, net `heat' per unit time is extracted from the cold and work reservoirs and dumped into the hot bath. That is, net energy is moved between the hot and cold baths \textit{against} the temperature gradient, with the assistance of the input heat coming from the work bath. Depending on the whether the useful effect is sought in the cold or the hot bath, we speak of a quantum absorption chiller or a quantum absorption heater. If on the contrary, the stationary rate of the heat-transforming cycle exceeds that of the cooling cycle, low-quality heat coming from the hot bath would be upgraded to high-temperature heat leaking into the work bath, only by means of the coupling to the cold bath. These are the two modes of operation of both classical and quantum absorption heat pumps \cite{gordon2000cool}.

At the steady state and given that each reservoir interacts \textit{locally} with its corresponding transition \cite{PhysRevE.87.042131}, it is intuitively clear that all three heat currents (work, hot and cold) must flow at the same rate, in order to keep the average energy asymptotically stationary \cite{PhysRev.156.343,1751-8121-44-49-492002}, i.e. one must have: $\dot{\mathcal{Q}}_\alpha=\omega_\alpha q$, where $\dot{\mathcal{Q}}_\alpha$ stands for the steady-state heat current flowing \textit{from} bath $\alpha\in\{w,h,c\}$ \textit{into} the system. Therefore
\begin{equation}
\left\vert\dot{\mathcal{Q_\alpha}}/\dot{\mathcal{Q}}_{\alpha'}\right\vert=\omega_\alpha/\omega_{\alpha'}.
\label{ideal}\end{equation}
Note that the frequency of the work transition $\ket{2}\leftrightarrow\ket{3}$ is just $\omega_w\equiv\omega_h-\omega_c$. We will take Eq.~\eqref{ideal} as the defining property of an \textit{ideal} heat pump. More generally, the asymptotic \textit{stationarity} of energy, translates into
\begin{equation}
\dot{\mathcal{Q}}_w+\dot{\mathcal{Q}}_h+\dot{\mathcal{Q}}_c=0,
\label{first_law}\end{equation}
which holds for both ideal and non-ideal devices. Since there is no work involved in the operation of the pump, Eq.~\eqref{first_law} is just a statement of the first law of thermodynamics. On another note, and always provided that the dissipation is \textit{sufficiently weak}, the positivity of the rate of irreversible entropy production \cite{spohn1978entropy,lindblad1976generators} makes it possible to write down the second law of thermodynamics in the form \cite{e15062100}
\begin{equation}
\frac{\dot{\mathcal{Q}}_w}{T_w}+\frac{\dot{\mathcal{Q}}_h}{T_h}+\frac{\dot{\mathcal{Q}}_c}{T_c}\leq 0.
\label{second_law}\end{equation}
Eqs.~\eqref{first_law} and \eqref{second_law} encode most of the relevant information about absorption heat pumps. Let us take for instance, an ideal device, assuming that it works in the chiller/heater mode, so that $\dot{\mathcal{Q}}_c,\dot{\mathcal{Q}}_w > 0$ and $\dot{\mathcal{Q}}_h<0$. Then, combining Eqs.~\eqref{ideal} and \eqref{second_law} one arrives to
\begin{equation}
0<\omega_c < \omega_{c,\max}\equiv\frac{(T_w-T_h)T_c}{(T_w-T_c)T_h}\omega_h,
\label{cooling_window}\end{equation}
which defines the \textit{cooling window}, that is, the region in parameter space in which cooling (and heating) is permitted by the second law. For $\omega_{c,\max} < \omega_c < \omega_h$, it is the heat-transforming cycle the one that dominates, thus rendering a quantum absorption heat transformer.

From the cooling point of view, the ratio of the incoming cold heat $\dot{\mathcal{Q}}_c$ and the energetic cost $\dot{\mathcal{Q}}_w$ of its processing, defines the efficiency or coefficient of performance (COP) $\varepsilon\equiv\dot{\mathcal{Q}}_c/\dot{\mathcal{Q}}_w$, which benchmarks the useful effect of an absorption chiller. Combining Eqs.~\eqref{first_law} and \eqref{second_law}, one generally finds
\begin{equation}
\varepsilon\leq\frac{(T_w-T_h)T_c}{(T_h-T_c)T_w}\equiv\varepsilon_C,
\label{carnot}\end{equation}
that is, that the maximum efficiency allowed is the corresponding Carnot COP $\varepsilon_C$ \cite{gordon2000cool}. It is easy to see that the Carnot efficiency is indeed saturated by an ideal fridge at $\omega_c\rightarrow\omega_{c,\max}$, where each transition equilibrates with its local bath and consequently, all heat currents vanish \cite{PhysRev.156.343}. In general, for a \textit{non-ideal} absorption chiller [i.e. a system not obeying Eq.~\eqref{ideal}], Eq.~\eqref{carnot} would be a strict inequality \cite{PhysRevE.87.042131}, while Eq.~\eqref{cooling_window} would cease to hold.
\newline

\begin{figure*}[htb]
\includegraphics[width=\textwidth]{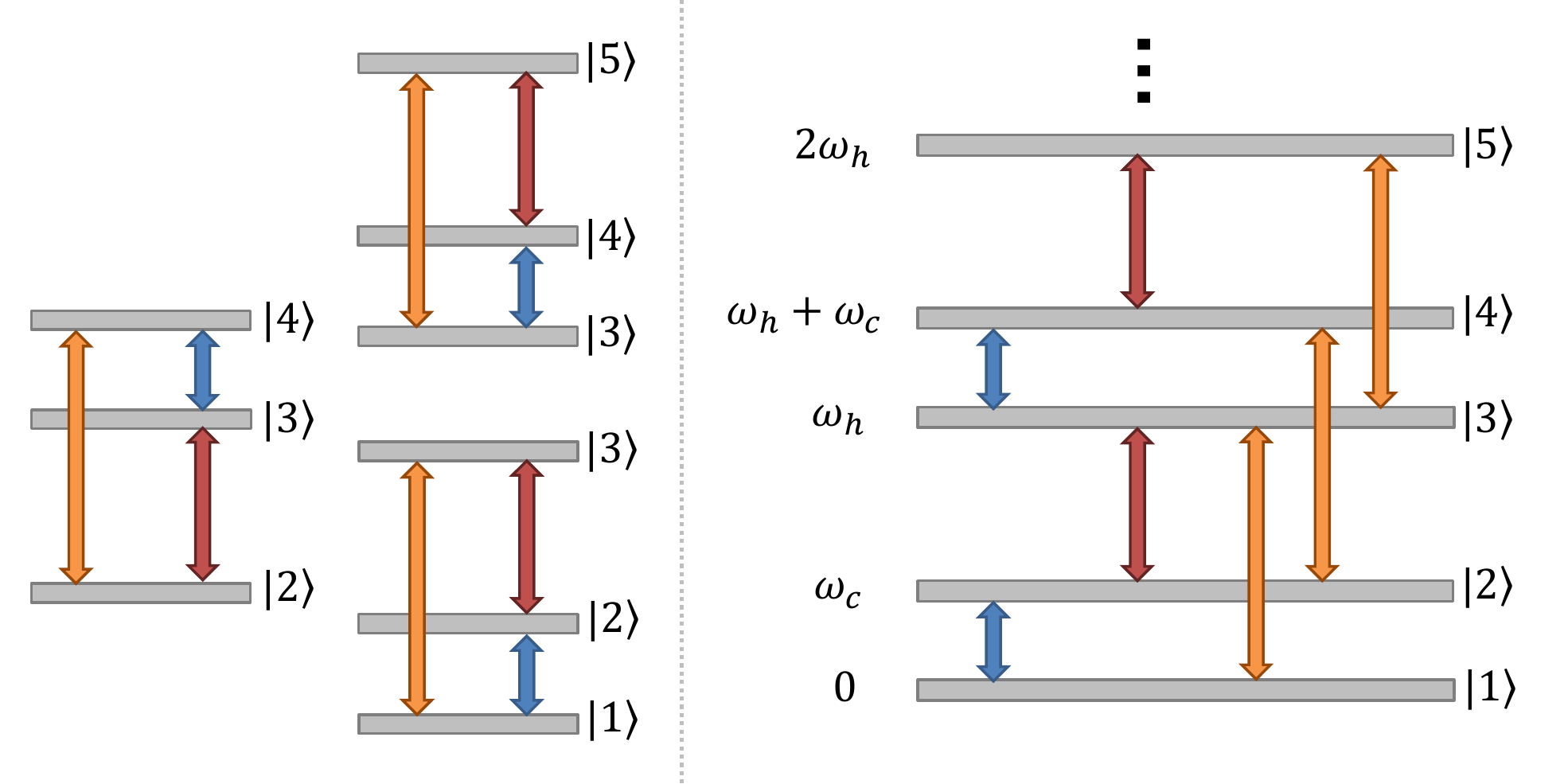}
\caption{On the right hand side, we show the detail of the first five levels of a generalized $N$-dimensional ideal quantum absorption heat pump. The colored arrows represent dissipative interaction between the corresponding transition and one of the three heat reservoirs: Blue stands for the cold bath, orange, for the hot bath, and red, for the high temperature work reservoir. Note that the energy difference between two consecutive levels keeps alternating between $\omega_c$ and $\omega_w = \omega_h - \omega_c$. On the left hand side, we show the \textit{qutrit breakup} of a 5-level fridge: It consists of three elementary three-level cycles operating in parallel. The first and the second share the work transition, while the second and the third, share a cold transition. The next elementary block to be added for a $6$-level heat pump, would share the work transition $\ket{4}\leftrightarrow\ket{5}$ with the preceding block, while $\ket{5}\leftrightarrow\ket{6}$ and $\ket{4}\leftrightarrow\ket{6}$ would be connected with the cold and hot baths respectively.}
\label{Fig1}
\end{figure*}

\section{Generalized $N$-level ideal heat pumps \label{system}}

An autonomous thermal device based on two non-interacting qubits was put forward in Ref.~\cite{PhysRevLett.108.070604}, which ultimately amounts to two elementary three-level cycles \textit{sharing} the work transition. It can be seen that it realizes a parallel double-stage ideal heat pump and that the partial heat currents from each stage just add up to the total \cite{1308.4174}. The overlap between them, however, makes their individual contributions smaller than what would be expectable from two independent three-level heat pumps. Once more, the correspondence between the classical and the quantum case is apparent \cite{gordon2000cool,srikhirin2001review,ziegler1993multi}, and the question naturally arises about the size scaling of the cooling power $\dot{\mathcal{Q}}_c$ and the COP $\varepsilon$ of multi-stage quantum absorption refrigerators. Does the size really matter in quantum absorption cooling?

\subsection{Microscopic model and master equation}

A generalized $N$-level ideal heat pump may be built by merging $N-2$ three-level systems, alternatively sharing a work or a cold transition. In Fig.~\ref{Fig1}, the case $N=5$ (triple-stage) is schematically illustrated (see figure caption for details). The total Hamiltonian of such $N$-level system reads
\begin{equation}
\hat H_N = \sum_{n=1}^{\lfloor N/2 \rfloor} \left[(n-1)~\omega_h+\omega_c\right]\ket{2n}+\sum_{n=1}^{\lceil N/2 \rceil-1} n~\omega_h\ket{2n+1},
\label{hamiltonianN}\end{equation}
where $\lfloor\cdot\rfloor$ and $\lceil\cdot\rceil$ stand for the floor and ceiling functions. Let us model the system-baths interactions with a term of the form $\hat H_{\text{S-B}}=\sum_\alpha\hat\Sigma^{(N)}_\alpha\otimes\hat{\mathcal{B}}_\alpha$, where
\begin{subequations}
\begin{align}
\hat\Sigma^{(N)}_{w} &=\sum\nolimits_{n=1}^{\lceil N/2 \rceil -1 } \ket{2n}\bra{2n+1}+\ket{2n+1}\bra{2n} \\
\hat\Sigma^{(N)}_{h} &=\sum\nolimits_{n=1}^{N-2} \ket{n}\bra{n+2}+\ket{n+2}\bra{n} \\
\hat\Sigma^{(N)}_{c} &=\sum\nolimits_{n=1}^{\lfloor N/2 \rfloor} \ket{2n-1}\bra{2n}+\ket{2n}\bra{2n-1}.
\end{align}
\label{sigma}\end{subequations}
The reservoirs may consist, as usual, of an infinite collection of uncoupled harmonic modes in three dimensions. We choose the bath coupling operators $\hat{\mathcal{B}}_\alpha$ to be
\begin{equation}
\hat{\mathcal{B}}_\alpha=\sum_\mu g_{\alpha\mu} \left(\hat b_{\alpha\mu} + {\hat b_{\alpha\mu}}^{\dagger}\right).
\label{baths}\end{equation}
With $\hat b_{\alpha\mu}$ and $\hat b_{\alpha\mu}^\dagger$, we denote the annihilation and creation operators for mode $\omega_\mu$ in the reservoir $\alpha$, and define the coupling constants as $g_{\alpha\mu}\equiv\sqrt{\gamma_\alpha\omega_\mu}$, which yields flat spectral densities ($J_\alpha(\omega)\propto\sum_\mu g_{\alpha\mu}^2/\omega_\mu$). The dissipation strengths $\gamma_\alpha$ are chosen so as to define the largest dynamical time-scale in the problem
\begin{equation}
\gamma_\alpha^{-1}\ggg\left\lbrace\vert\omega_\alpha-\omega_{\alpha'}\vert^{-1},\frac{\hbar}{k_B T_\alpha}\right\rbrace.
\label{markov_rwa}\end{equation}
This allows us to perform the Born-Markov and secular approximations in deriving a master equation for the system \cite{breuer2002theory}. Under the further assumption of a preparation with totally uncorrelated system and equilibrium reservoirs, one readily obtains
\begin{equation}
\frac{d}{dt}\hat\varrho_N(t)=\left(\mathcal{L}^{(N)}_w+\mathcal{L}^{(N)}_h+\mathcal{L}^{(N)}_c\right)\hat\varrho_N(t),
\label{master_equation}\end{equation}
where $\hat\varrho_N(t)$ is the state of the system in the interaction picture and the \textit{dissipators} $\mathcal{L}^{(N)}_\alpha$ are given by
\begin{multline}
\mathcal{L}^{(N)}_\alpha\hat\varrho=\Gamma_{\alpha,\omega_\alpha}\left(\hat\sigma_{N,\alpha}^{-}\hat\varrho\hat\sigma_{N,\alpha}^{+}-\frac{1}{2}\hat\sigma_{N,\alpha}^{+}\hat\sigma_{N,\alpha}^{-}\hat\varrho-\frac{1}{2}\hat\varrho\hat\sigma_{N,\alpha}^{+}\hat\sigma_{N,\alpha}^{-}\right) +\\
\Gamma_{\alpha,-\omega_\alpha}\left(\hat\sigma_{N,\alpha}^{+}\hat\varrho\hat\sigma_{N,\alpha}^{-}-\frac{1}{2}\hat\sigma_{N,\alpha}^{-}\hat\sigma_{N,\alpha}^{+}\hat\varrho-\frac{1}{2}\hat\varrho\hat\sigma_{N,\alpha}^{-}\hat\sigma_{N,\alpha}^{+}\right).
\label{dissipator}\end{multline}
Here, the positive \textit{decay rates} $\Gamma_{\alpha,\omega_\alpha}$ are the diagonal elements of the spectral correlation tensor \cite{breuer2002theory}, which for a three-dimensional free bosonic field in thermal equilibrium are given by $\Gamma_{\alpha,\omega_\alpha}\propto\omega^3[1+n_\alpha(\omega)]>0$, with $n_\alpha(\omega)=(\exp{\hbar\omega/k_B T_\alpha-1})^{-1}$. The mutually adjoint \textit{jump} operators $\hat\sigma_{N,\alpha}^{+}=(\hat\sigma_{N,\alpha}^{-})^\dagger$ are just
\begin{subequations}
\begin{align}
\hat\sigma_{N,w}^{-} &= \sum\nolimits_{n=1}^{\lceil N/2 \rceil -1 } \ket{2n}\bra{2n+1} \\
\hat\sigma_{N,h}^{-} &= \sum\nolimits_{n=1}^{N-2} \ket{n}\bra{n+2} \\
\hat\sigma_{N,c}^{-} &= \sum\nolimits_{n=1}^{\lfloor N/2 \rfloor} \ket{2n-1}\bra{2n}
\label{jump_operators}\end{align}
\end{subequations}
The dissipators in Eq.~\eqref{dissipator} are in the standard Lindblad form \cite{lindblad1976generators}, which guarantees that the reduced dynamics of the system is completely positive and trace-preserving. Each of them individually, generates a fully contractive dynamics of the system towards its `local' equilibrium states $\propto\exp\{-\hat H_N/k_BT_\alpha\}$ \cite{spohn1978entropy}, which in turn, implies the second law of thermodynamics as expressed by Eq.~\eqref{second_law} \cite{e15062100,PhysRevE.85.061126}.

\subsection{Stationary heat currents}

In the light of Eqs.~\eqref{master_equation} and \eqref{dissipator}, our definition of steady-state heat current from Sec.~\ref{review} translates into $\dot{\mathcal{Q}}^{(N)}_\alpha\equiv\text{tr}\{\hat H_N\mathcal{L}^{(N)}_\alpha\hat\varrho_N(\infty)\}$ \cite{PhysRevE.64.056130}, where the asymptotic state $\hat\varrho_N(\infty)$ results from equating to zero the right-hand side of Eq.~\eqref{master_equation}. 

In the simplest case of $N=3$, one easily finds
\begin{subequations}
\begin{align}
\dot{\mathcal{Q}}^{(3)}_{w} &=\omega_w\bra{3}\mathcal{L}^{(3)}_{w}\hat\varrho_3(\infty)\ket{3} \label{heat_currents_3_w}\\
\dot{\mathcal{Q}}^{(3)}_{h} &=\omega_h\bra{3}\mathcal{L}^{(3)}_{h}\hat\varrho_3(\infty)\ket{3} \label{heat_currents_3_h}\\
\dot{\mathcal{Q}}^{(3)}_{c} &=\omega_c\bra{2}\mathcal{L}^{(3)}_{c}\hat\varrho_3(\infty)\ket{2} \label{heat_currents_3_c} .
\end{align}
\label{heat_currents_3}\end{subequations}
The addition of a second stage to the heat pump introduces the further cold and hot transitions $\ket{3}\stackrel{c}\leftrightarrow\ket{4}$ and $\ket{2}\stackrel{h}\leftrightarrow\ket{4}$ that contribute to the total steady-state heat currents to yield
\begin{subequations}
\begin{align}
\dot{\mathcal{Q}}^{(4)}_{w} &=\omega_w\bra{3}\mathcal{L}^{(4)}_{w}\hat\varrho_4(\infty)\ket{3} \label{heat_currents_4_w}\\
\dot{\mathcal{Q}}^{(4)}_{h} &=\omega_h\left(\bra{3}\mathcal{L}^{(4)}_{h}\hat\varrho_4(\infty)\ket{3}+\bra{4}\mathcal{L}^{(4)}_{h}\hat\varrho_4(\infty)\ket{4}\right) \label{heat_currents_4_h}\\
\dot{\mathcal{Q}}^{(4)}_{c} &=\omega_c\left(\bra{2}\mathcal{L}^{(4)}_{c}\hat\varrho_4(\infty)\ket{2}+\bra{4}\mathcal{L}^{(4)}_{c}\hat\varrho_4(\infty)\ket{4}\right) \label{heat_currents_4_c}.
\end{align}
\label{heat_currents_3}\end{subequations}
In general, for arbitrary $N$, one would be left with
\begin{subequations}
\begin{align}
\dot{\mathcal{Q}}^{(N)}_{w} &=\omega_w\sum\nolimits_{n=1}^{\lceil N/2 \rceil-1} \bra{2n+1}\mathcal{L}^{(N)}_w\hat\varrho(\infty)\ket{2n+1} \label{heat_currents_N_w}\\
\dot{\mathcal{Q}}^{(N)}_{h} &=\omega_h\sum\nolimits_{n=3}^{N} \left(\lceil n/2 \rceil-1\right)\bra{n}\mathcal{L}^{(N)}_h\hat\varrho(\infty)\ket{n} \label{heat_currents_N_h}\\
\dot{\mathcal{Q}}^{(N)}_{c} &=\omega_c\sum\nolimits_{n=1}^{\lfloor N/2 \rfloor} \bra{2n}\mathcal{L}^{(N)}_w\hat\varrho(\infty)\ket{2n} \label{heat_currents_N_c},
\end{align}
\label{heat_currents_N}\end{subequations}
which provides, together with Eqs.~\eqref{sigma} and \eqref{dissipator}, closed formulas for the evaluation of the stationary heat currents. From now on, we shall focus on the chiller operation mode. At first glance, one could intuitively expect a different scaling of e.g. $\vert\dot{\mathcal{Q}}^{(N)}_c\vert$ for even and odd $N$: The addition of an even level to the heat pump provides it with an extra cold transition, so that a new term contributes to the cold current in Eq.~\eqref{heat_currents_N_c} while, when the added level is odd, the new contribution appears instead in the work current of Eq.~\eqref{heat_currents_N_w}. It must be noted, however, that each of the already contributing terms will always decrease as $N$ grows, e.g. $\vert\bra{2}\mathcal{L}^{(3)}_{c}\hat\varrho_3(\infty)\ket{2}\vert<\vert\bra{2}\mathcal{L}^{(4)}_{c}\hat\varrho_4(\infty)\ket{2}\vert$ when going from Eq.~\eqref{heat_currents_3_c} to Eq.~\eqref{heat_currents_4_c}.

It can be seen by numerical inspection of Eqs.~\eqref{heat_currents_N} that increasing the number of stages from even to odd is indeed detrimental for \emph{all} currents, i.e. $\vert\dot{\mathcal{Q}}_\alpha^{(N+1)}\vert\leq\vert\dot{\mathcal{Q}}_\alpha^{(N)}\vert$, when $N$ is even. Still for arbitrary $N$, one always finds $\vert\dot{\mathcal{Q}}_\alpha^{(N+2)}\vert-\vert\dot{\mathcal{Q}}_\alpha^{(N)}\vert\geq 0$, though the enhancement decreases as the system scales up.

For practical purposes, one would wish to optimize the incoming cold heat current to cool as \textit{quickly} as possible. Given a set of heat baths $T_\alpha$ and dissipation rates $\gamma_\alpha$, one has to find the cold frequency $0\leq\omega_c\leq\omega_{c,\max}$ that renders the maximum cooling power $\dot{\mathcal{Q}}_{c,\max}$ for every fixed $\omega_h$. The corresponding COP $\varepsilon_*$ is a sensible figure of merit for cooling and, as rigorously proven in \cite{1308.4174}, it is tightly upper bounded by $\frac{d}{d+1}\varepsilon_C$ both in the single and double-stage cases, where $d$ stands for the \emph{spatial} dimensionality of the cold bath, i.e. in our case $d=3$. Once parameter optimization has taken place at the single-stage level, we shall be concerned with the potential enhancement in both $\varepsilon_*$ and $\dot{\mathcal{Q}}_{c,\max}$ as the number of stages, and thus the system's complexity, grows. 
\newline

\begin{figure*}[htb!]
	\subfigure{\label{Fig2a}
	\includegraphics[scale=0.55]{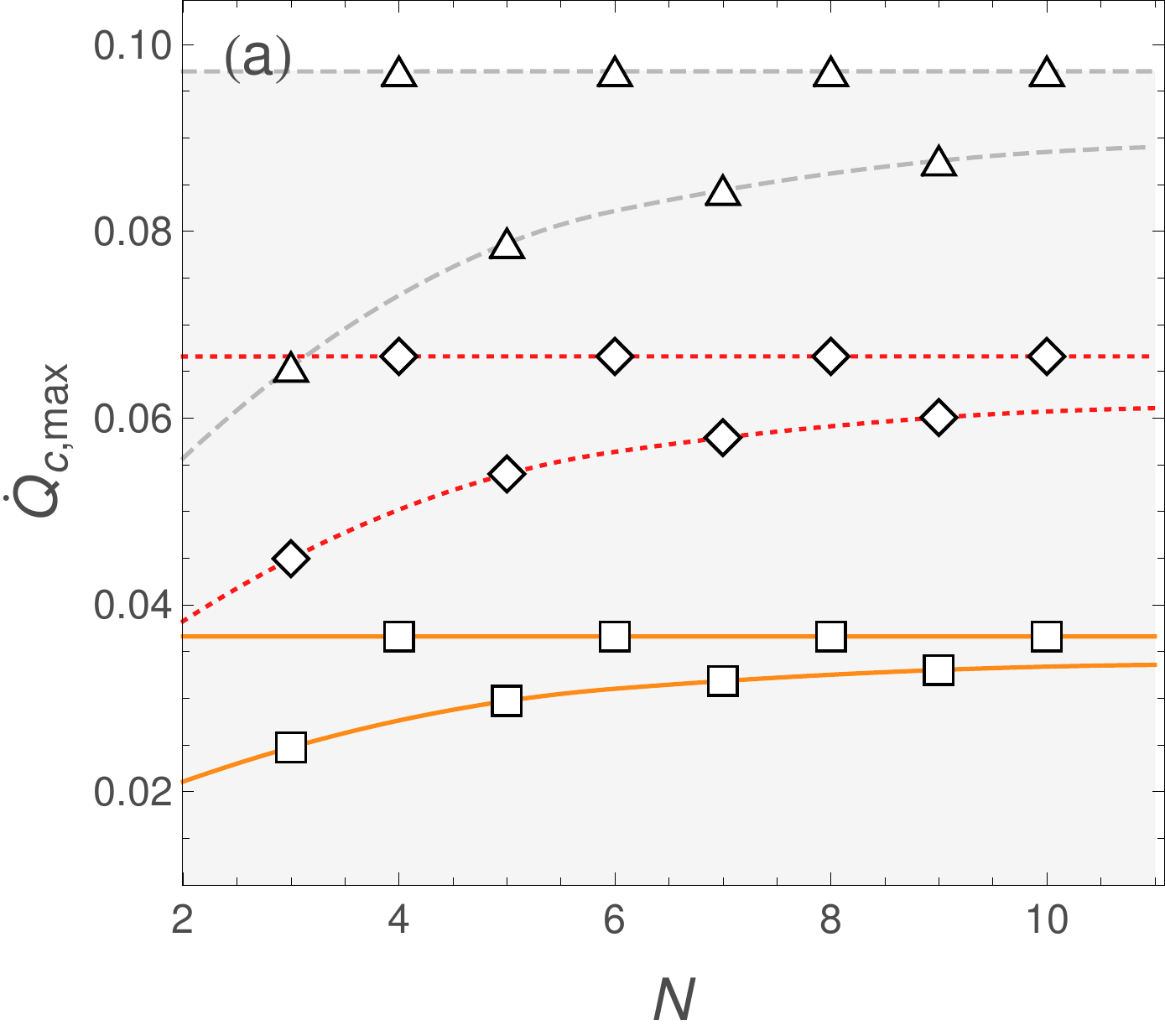}}
	\subfigure{\label{Fig2b}
	\includegraphics[scale=0.54]{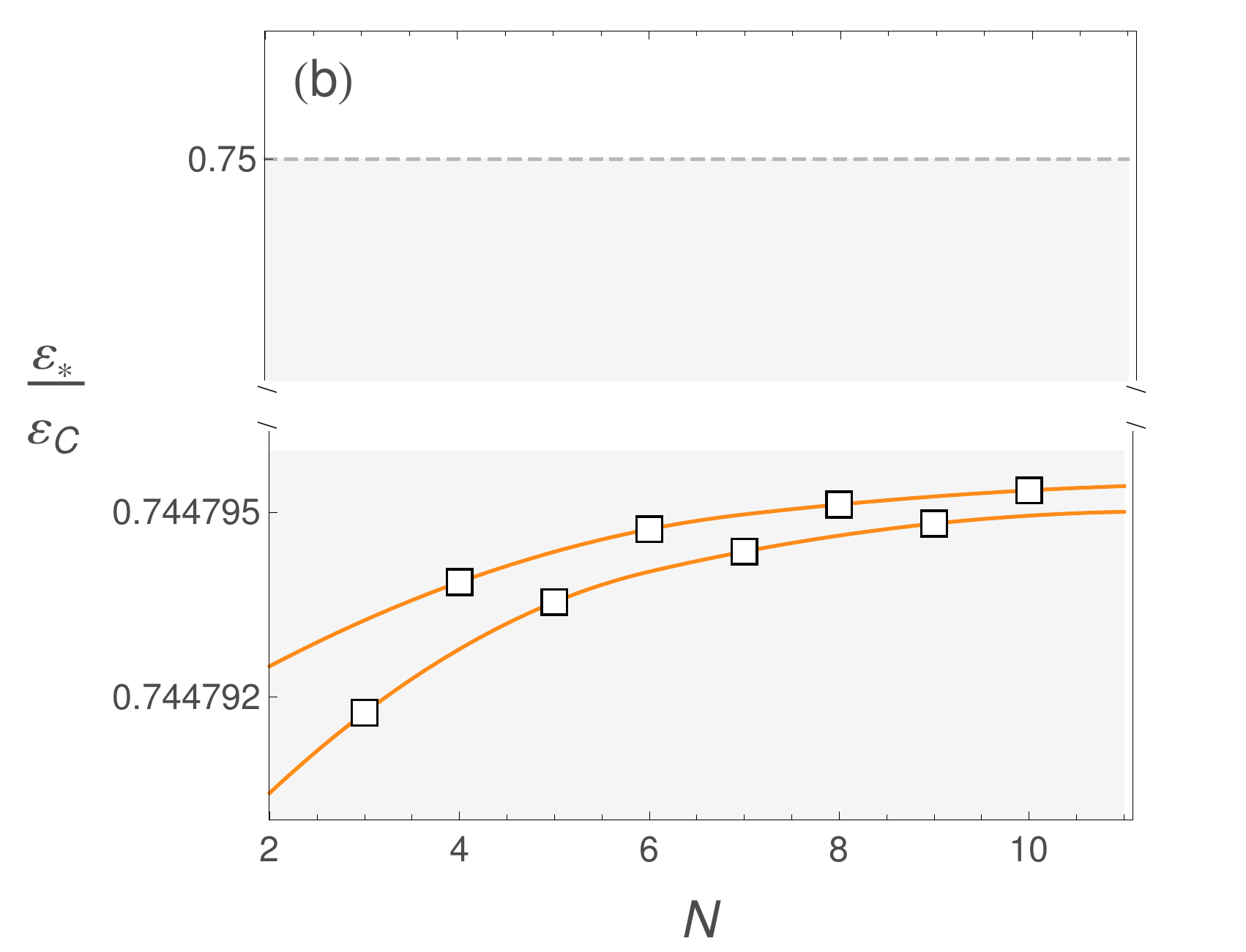}}
\caption{(a) (Squares) Maximum cooling power as a function of $N$ for generalized multi-stage heat pumps in the chiller configuration, with parameters: $T_w=7.1\times 10^3$, $T_h=1.57\times10^3$, $T_c=54.25$, $\gamma_w=3.5\times10^{-3}$, $\gamma_h=5.1\times10^{-3}$, $\gamma_c=8.8\times10^{-3}$ and $\omega_h=102.6$ ($\hbar=k_B=1$). (Rhombs) Maximum cooling power for increasing $N$ when the modes of the work reservoir are all squeezed by $7$ dB, which effectively raises the work temperature to $T_w\simeq 1.8\times10^4$. (Triangles) Maximum cooling power versus $N$ for reversed multi-stage Carnot engines, i.e. heat pumps with \textit{saturated} work transitions. The solid orange, dotted red and dashed blue curves are merely a guide to the eye, highlighting the different size scaling of devices with even and odd number of parallel cycles. The domain reachable by absorption chillers is depicted in shaded gray. (b) Efficiency at maximum power as a function of $N$, for the same parameters as in (a). The shaded gray region correspond the the domain of $\varepsilon_*$ allowed by a bosonic cold bath in three dimensions with flat spectral density \cite{1308.4174}.}
\label{Fig2}
\end{figure*}

\section{Results and discussion \label{results}}

\subsection{Maximum power and COP at maximum power}

The dependence of the maximum cooling power on the number of stages is illustrated in Fig.~\ref{Fig2a}. The temperatures $T_\alpha$, dissipation rates $\gamma_\alpha$ and the hot frequency $\omega_h$ were chosen at random so as to fix the time scales for thermal fluctuations, dissipation and system's free evolution. Then, the optimization in $\omega_c$ was carried out within the cooling window. 

As could be expected from the even/odd oscillations of the stationary heat currents, $\dot{\mathcal{Q}}_{c,\max}(N)$ scales differently for even and odd $N$. While in the even case, the cooling power barely increases, odd heat pumps significantly benefit from scaling up. One also sees that any even configuration outperforms all the odd ones. In contrast, the corresponding COP at maximum power $\varepsilon_*$ proves to be roughly size-independent, as shown in Fig.~\ref{Fig2b}. Consequently, the double-stage absorption heat pump \cite{PhysRevLett.108.070604} tuned to operate at maximum power, appears as the best possible choice in terms of system optimization. Our extensive numerics show that these properties are completely general in the range of application of Eqs.~\eqref{master_equation} and \eqref{dissipator}.

It is always possible to enhance both the cooling power and the corresponding COP, by suitably engineering the spectrum of the environments \cite{PhysRevE.86.051105,1303.6558v1,1308.5935v1}. Perhaps the simplest thing to try is squeezing the modes of the work bath, which effectively raises its temperature $T_w$ and boosts the output power of the heat pump \cite{1308.4174}. The Carnot bound $\varepsilon_C$ is also effectively increased, which allows for larger efficiencies at maximum power. The effect of reservoir squeezing in the cooling power is also illustrated in Fig.~\ref{Fig2a} for a squeezing of $7$ dB or $r\simeq 0.8$ (see figure caption for details). As $T_w$ increases, the corresponding contact transitions tend to saturate. In the limit of $T_w\rightarrow\infty$, one realizes generalized multi-stage Carnot engines running in reverse \cite{PhysRev.156.343} and thus, abandons the realm of absorption chillers. The limiting cooling powers are also depicted in Fig.~\ref{Fig2a}.

\subsection{Bound on the COP at maximum power}

The quest for tight bounds to the COP at maximum power for heat engines and refrigerators has attracted a lot of interest over the last decades, both in the classical \cite{curzon1975efficiency,PhysRevLett.105.150603,PhysRevLett.78.3241} and quantum \cite{PhysRevE.81.051129,abe2011power,PhysRevE.87.042131,wang2012finite}
scenarios. By making minimal assumptions about the dominating sources of irreversibility, the aim is to derive practical bounds, as widely applicable as possible, to benchmark the operation of realistic devices. The paradigmatic example is the Curzon-Ahlborn bound \cite{curzon1975efficiency}: Although obtained for a specific phenomenological modeling of the sources of irreversibility, i.e. the \textit{endoreversible} approximation, it emerges naturally as a limiting case in different instances of heat engine \cite{esposito2009universality,PhysRevLett.105.150603,wang2012finite} and describes well the performance of actual power plants.

\begin{figure*}[htb!]
	\subfigure{\label{Fig3a}
	\includegraphics[scale=0.50]{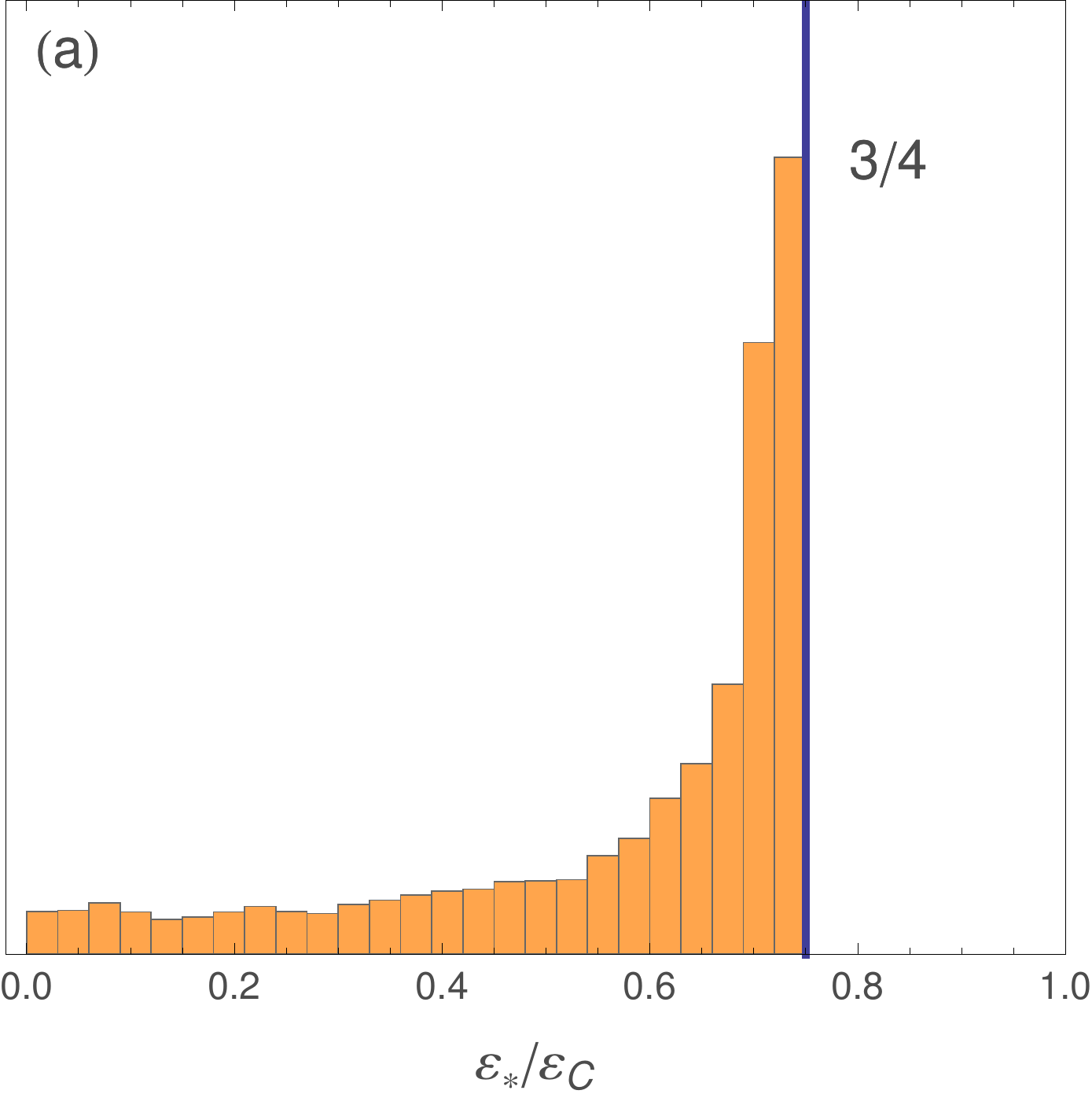}}
	\subfigure{\label{Fig3b}
	\includegraphics[scale=0.53]{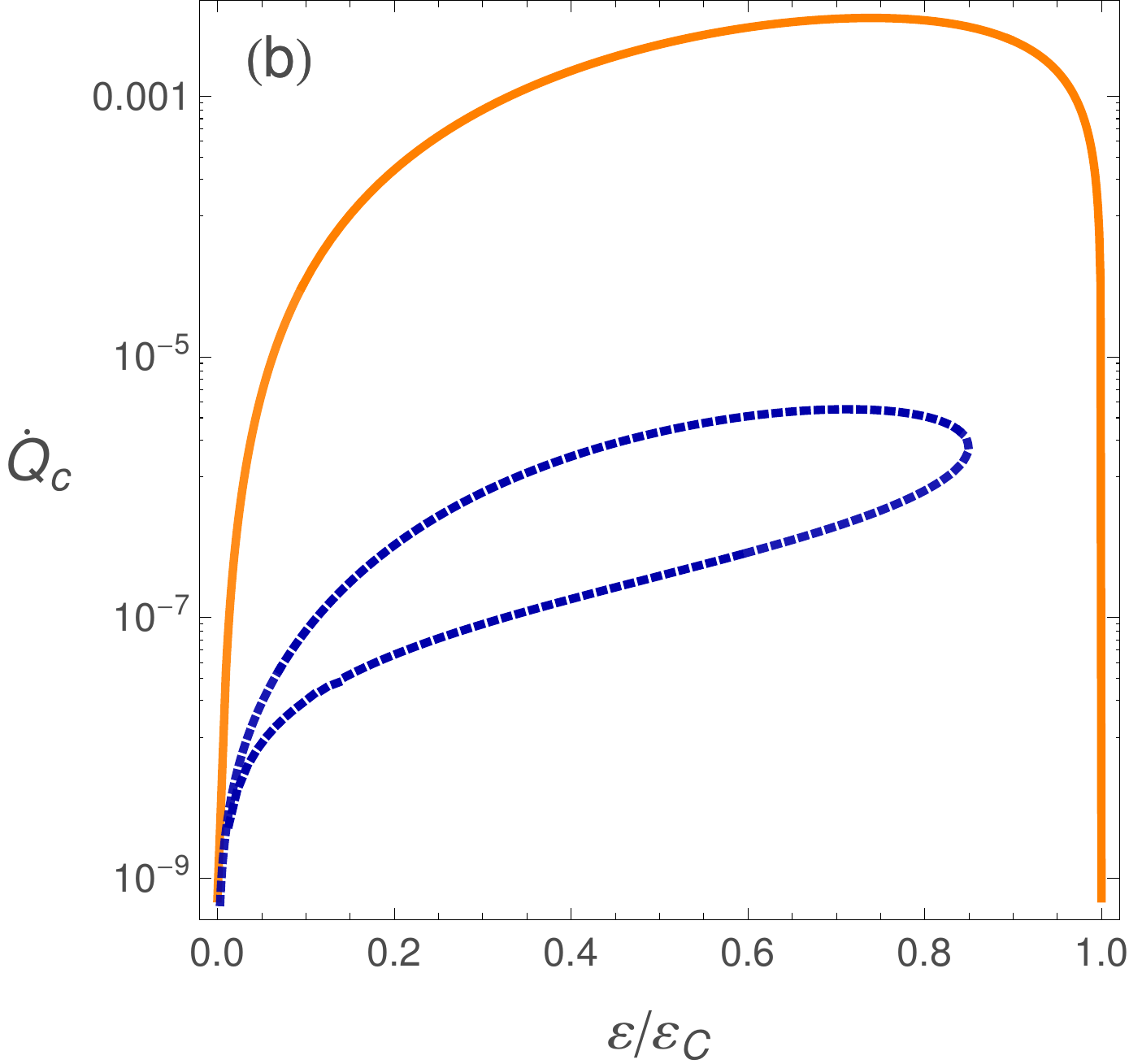}}
\caption{(a) Histogram of the COP at maximum power for $10^5$ multi-stage absorption fridges with $\omega_h$, $T_\alpha$, $\gamma_\alpha$ and $N$ chosen at random. The blue vertical line marks the $\frac{3}{4}\varepsilon_C$ ultimate bound on $\varepsilon_*$ \cite{1308.4174}. (b) Cooling power versus normalized efficiency for the three-qubit non-ideal fridge of \cite{PhysRevLett.105.130401} (solid orange), and for an eight-level ideal chiller (dotted blue) for $\omega_w=60$, $T_w=130$, $T_h=60$, $T_c=5$ and $\gamma_\alpha=10^{-3}$. The three-body interaction strength between the qubits in the non-ideal model was set to $g=0.1$. Again, natural units are assumed.}
\label{Fig3}
\end{figure*}

The bound $\varepsilon_*<\frac{3}{4}\varepsilon_C$, derived from first principles for the three-level maser and the four-level double stage chiller in \cite{1308.4174}, has been also seen to hold \cite{PhysRevE.87.042131} in the non-ideal eight-level refrigerator model of \cite{PhysRevLett.105.130401}, and even succeeds in limiting from above the performance of classical endoreversible absorption fridges \cite{PhysRevA.39.4140}. In order to establish whether it continues to hold for multi-stage ideal absorption refrigerators beyond the cases $N=3$ and $N=4$, we performed global numerical optimization of $\varepsilon_*$ over $\omega_h$, $T_\alpha$, $\gamma_\alpha$ and $N\in [3,10]$. An histogram on the resulting COP at maximum power for $10^5$ randomly drawn refrigerators is presented in Fig.~\ref{Fig3a}, clearly showing that the bound holds \textit{tightly} regardless of the size of the system.

\subsection{Ideal vs. non-ideal absorption fridges}

A lot of interest has been generated by the non-ideal eight-level (three-qubit) absorption fridge, first introduced in \cite{PhysRevLett.105.130401}, including experimental proposals for its implementation on super-conducting qubits \cite{0295-5075_97_4_40003} and quantum dots \cite{PhysRevLett.110.256801}. It is legitimate to ask how does it compare, in terms of absolute cooling power, with its ideal eight-level counterpart. Interestingly, all ideal multi-stage fridges (even the three-level maser), can be seen to largely outperform the eight level non-ideal chiller, typically by several orders of magnitude. 

In Fig.~\ref{Fig3b} this is illustrated by plotting together the performance characteristics of both the non-ideal three-qubit and the ideal `sextuple-stage' ($N=8$) chiller, for the same choice of parameters. A closed curve in the $\dot{\mathcal{Q}}_c$ -- $\varepsilon$ plane is indicative of irreversibility and prevents the refrigerator from ever realizing the Carnot efficiency. The COP at maximum power is roughly the same in both cases (close to its ultimate $\frac34\varepsilon_C$ bound), but the power $\dot{\mathcal{Q}}_{c,\max}$ of the ideal fridge exceeds that of the non-ideal one by over three orders of magnitude.

This suggests that any realistic application of absorption technologies to quantum cooling, should rely upon physical implementations of ideal (multi-stage) heat pumps, out of which the double-stage design \cite{PhysRevLett.108.070604} stands out with its optimal balance between performance and simplicity.

\section{Conclusions \label{conclusions}}

We have studied the size scaling of the power and the efficiency of parallel multi-stage quantum absorption heat pumps. When working in the chiller configuration, heat pumps comprised of an odd and even number stages, scale differently in terms of their maximum achievable cooling power: Devices with even $N$ always provide a larger power, which is almost size-independent, while in heat pumps with odd $N$, the maximum power increases with the number of stages, quickly saturating to a constant asymptotic value. Contrarily, the efficiency at maximum power does not significantly scale with $N$ and, upon global optimization over all free parameters, it saturates to the same model-independent bound, regardless of the size of the system.

Even if the heat currents were formally given in Eqs.~\eqref{dissipator} and \eqref{heat_currents_N}, the explicit formulas for any $N$ are quite involved and nothing but insightful. The whole analysis is therefore based on extensive numerics. It must be acknowledged as well, that the working substance of an ideal multi-stage absorption refrigerator is generally a rather artificial quantum system. Interestingly enough, the case of $N=4$, that appears as the optimal compromise between performance and complexity, also turns out to correspond to the two-qubit model of \cite{PhysRevLett.108.070604}. This suggests that it should be considered for practical applications of absorption cooling to quantum technologies.
 
The double-stage fridge (more generally, \textit{any} ideal heat pump) was seen to outperform by several orders of magnitude the maximum cooling power delivered by the only available (non-ideal) alternative model: The three-qubit refrigerator \cite{PhysRevLett.105.130401}. Arrangements of superconducting qubits \cite{0295-5075_97_4_40003} or coupled quantum dots \cite{PhysRevLett.110.256801}, similar to those proposed to realize the three-qubit device, could also be good candidates to support the more promising double-stage absorption chillers. 

Further enhancement of both the power and the efficiency can always be achieved by applying quantum reservoir engineering techniques \cite{PhysRevA.57.548,shahmoon2013engineering} meant to render exploitable nonequilibrium environmental fluctuations, thus raising absorption technologies to the level of the best compression-based quantum thermodynamic cycles.
\newline

\section*{Acknowledgements}
The author is grateful to G. Adesso, J. P. Palao, D. Alonso, K. Hovhannisyan, P. Skrzypczyk, N. Brunner and P. Ramos-Garc\'{i}a for fruitful discussions and useful comments. This work was supported by the COST Action MP1006 and by the Canary Islands Government through the ACIISI fellowships (85\% co funded by European Social Fund).

\bibliographystyle{apsrev}
%\bibliography{/home/luis/Desktop/Current/10/References}

\end{document}